\RequirePackage{fix-cm}
\documentclass{svjour3}                     % onecolumn (standard format)
\smartqed  % flush right qed marks, e.g. at end of proof
\usepackage[dvipdfmx]{graphicx} 
\usepackage{amsmath}
\usepackage{braket}
\usepackage{bm,color}
\usepackage{here}
\usepackage{subcaption}
\usepackage{booktabs}
\usepackage{multirow}
\begin{document}

\title{Three-$\alpha$ configurations in the $0^{+}$ states of $^{12}\mathrm{C}$
  \thanks{
This work was in part supported by JSPS KAKENHI Grants
Nos.\ 18K03635 and 19H05140.}
%Grants or other notes
}


\author{H. Moriya         \and
        W. Horiuchi       \and 
        J. Casal \and \\
        L. Fortunato   %etc.
}


\institute{H. Moriya and W. Horiuchi \at
              Department of Physics, Hokkaido University, Sapporo, 060-0810 Japan \\
              \email{moriya@nucl.sci.hokudai.ac.jp, whoriuchi@nucl.sci.hokudai.ac.jp}           %  \\
          %  \and
          %  S. Author \at
              %     second address
  \and J. Casal and L. Fortunato  \at
  Dipartimento di Fisica e Astronomia ``G. Galilei'', Universit\`a degli
  Studi di Padova and   Istituto Nazionale di Fisica Nucleare-Sezione di Padova,
  via Marzolo 8, Padova I-35131, Italy
%              \email{ }           %  \\              
}
\date{Received: date / Accepted: date}

\maketitle

\begin{abstract}
  Geometric configurations of three-$\alpha$ particles in
  the ground- and first-excited $J^\pi=0^+$ states of $^{12}$C
  are discussed within two types of $\alpha$-cluster models
  which treat the Pauli principle differently.
Though there are some quantitative differences especially in
the internal region of the wave functions,
equilateral triangle configurations
are dominant in the ground state,
while in the first excited $0^+$ state isosceles triangle configurations
dominate, originating from $^8{\rm Be}+\alpha$ configurations.
% Insert your abstract here. Include keywords, PACS and mathematical
% subject classification numbers as needed.
\end{abstract}

\vspace{-0.3cm}
\section{Introduction}
\label{intro}
An $\alpha$ ($^4$He nucleus) clustering phenomenon
is one of the most important aspects of nuclear physics.
In particular, the first excited $J^{\pi}=0^{+}$ state,
$0_{2}^{+}$, the so-called Hoyle state
has attracted attention as it plays a crucial role
in the nucleosynthesis~\cite{h54} and
in a context of the Bose-Einstein condensed (BEC) state~\cite{THSR}.
Motivated by the recent progress in the structure study
based on algebraic cluster models~\cite{f19,vcfl20},
here we discuss geometric configurations
of $\alpha$ particles in the $0_1^+$ and $0_2^+$ states of $^{12}$C
using precise three-$\alpha$ wave functions.
In this paper, we treat the $\alpha$ particle as a point particle.
To incorporate the Pauli principle between two $\alpha$ particles,
two types of potential models are employed:
the orthogonality condition model~\cite{OCM}
and the shallow potential model~\cite{ab65}.
In the orthogonality condition model,
the Pauli principle is taken into account
by imposing the orthogonality condition on the Pauli forbidden states
for the relative motion between $\alpha$ particles.
The resulting radial wave function is pushed out,
exhibiting some nodes in the internal region.
In the shallow potential model, instead of using the orthogonality condition,
a repulsive component at short distances is introduced
to push the internal wave function out.
It is known that those two phase-shift equivalent potentials give
quite different results for some observables~\cite{Pinilla11,Arai18}.
We will discuss how these model difference affects
the geometric configurations in the low-lying $0^+$ states of $^{12}$C.

\vspace{-0.3cm}
\section{Results and discussion}
\label{sec:2}

We start from a standard three-$\alpha$ Hamiltonian
that consists of the kinetic, two-$\alpha$, and three-$\alpha$ potential
terms. The Coulomb potential is included.
For the OCM, we adopt the same Hamiltonian used in Ref.~\cite{kk05,kk07}.
This potential is deep enough to accommodate three redundant forbidden states.
To exclude the Pauli forbidden states between $\alpha$ particles,
we impose the orthogonality condition practically
by adding a pseudo-potential to the Hamiltonian~\cite{Kukulin78}.
As a shallow potential model,
we employ a slightly modified version
of the Ali-Bodmer potential (AB)~\cite{ab65}
(Set a$^\prime$~\cite{Fedorov96}),
which is $l$-dependent and widely used to describe
the Hoyle state~\cite{i13,i14,nnt13}. 

A three-$\alpha$ wave function $\Psi_{3\alpha}$ is described as 
a superposition of symmetrized correlated Gaussians
with the aid of the stochastic variational method~\cite{vs95,SVM}.
Details of the basis optimization can be found in Ref.~\cite{LNP20}. 
Also see, e.g., Refs.~\cite{Suzuki08,Mitroy13,Suzuki17},
for many examples showing the power of this approach.
For the OCM, the calculated energy and
the root-mean-square (rms) point-$\alpha$ radius
of the $0_{1}^{+}$ ($0_{2}^{+}$) state
are $-7.13$ (0.78) MeV and $1.73$ (4.58) fm, respectively.
For the AB, the values are $-6.67$ (0.38) MeV and 1.84 (3.39) fm,
respectively. These results are consistent with the ones
obtained in Refs.~\cite{ofkh13,i13}, respectively.

\begin{figure}[ht]
    \includegraphics[width=0.5\hsize]{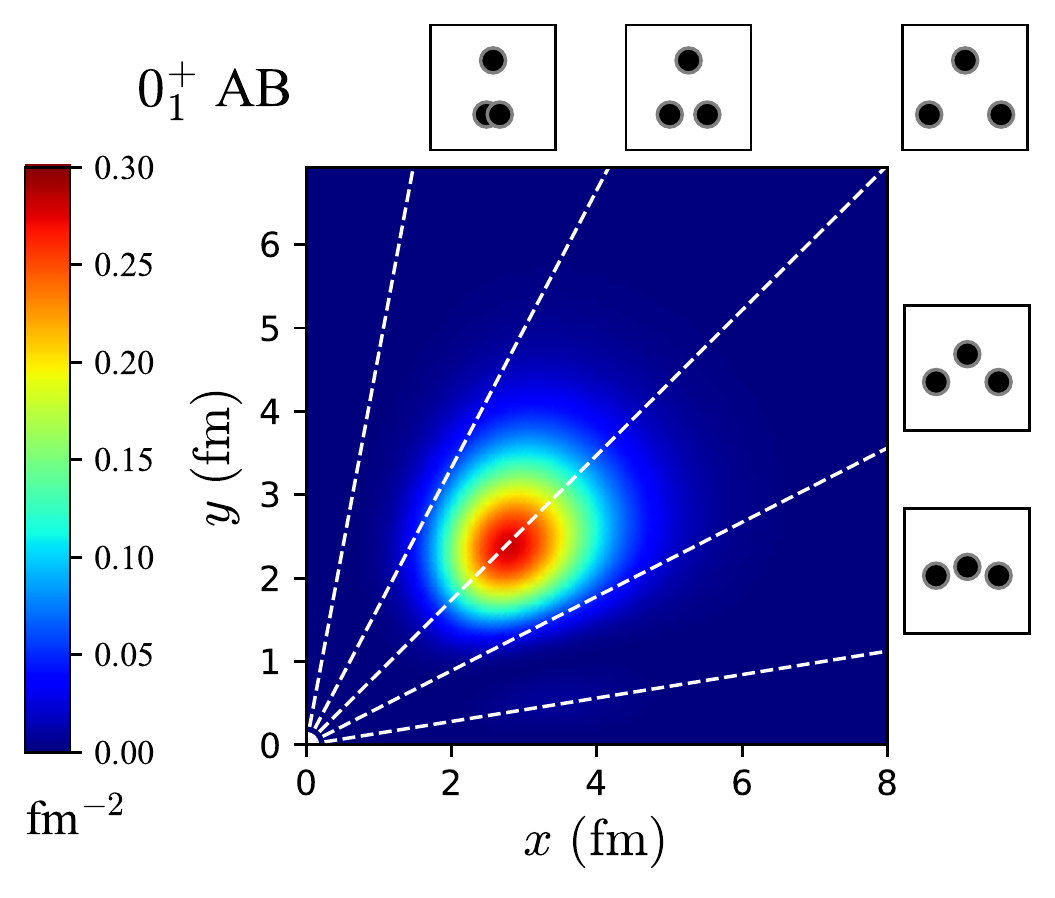}
    \includegraphics[width=0.5\hsize]{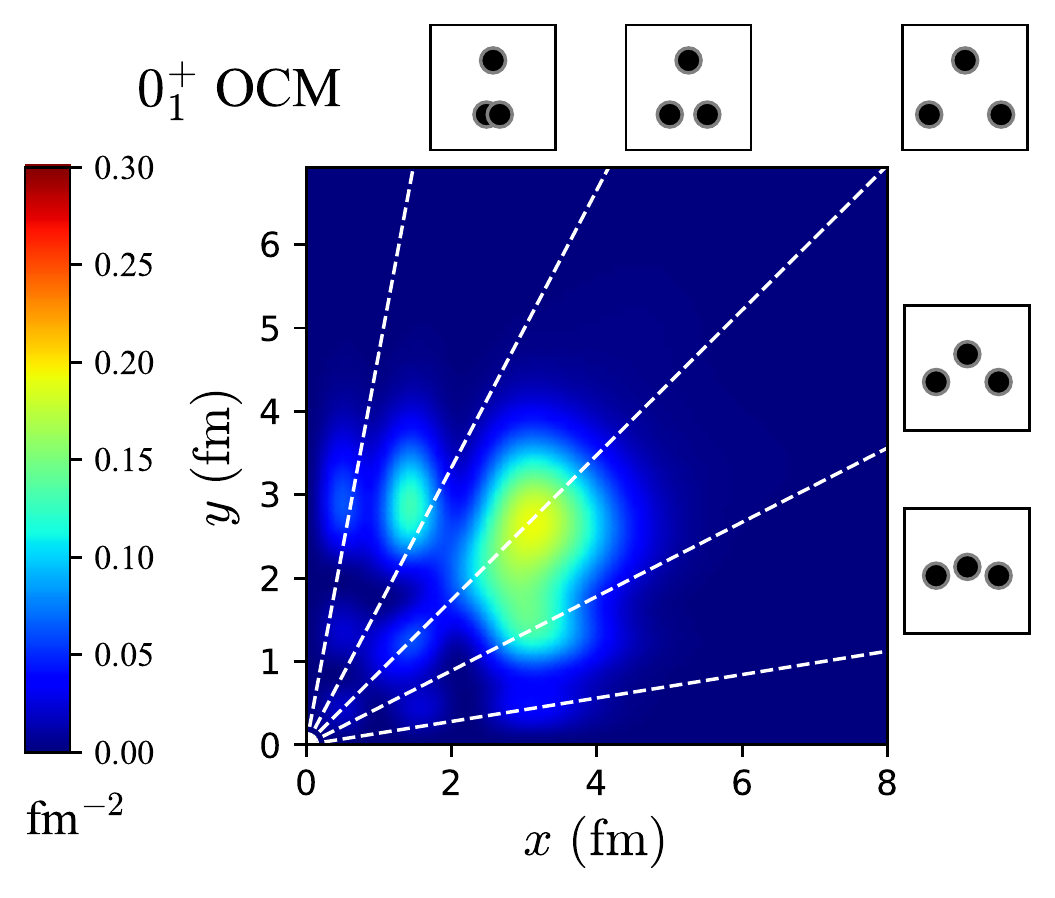}
    \includegraphics[width=0.5\hsize]{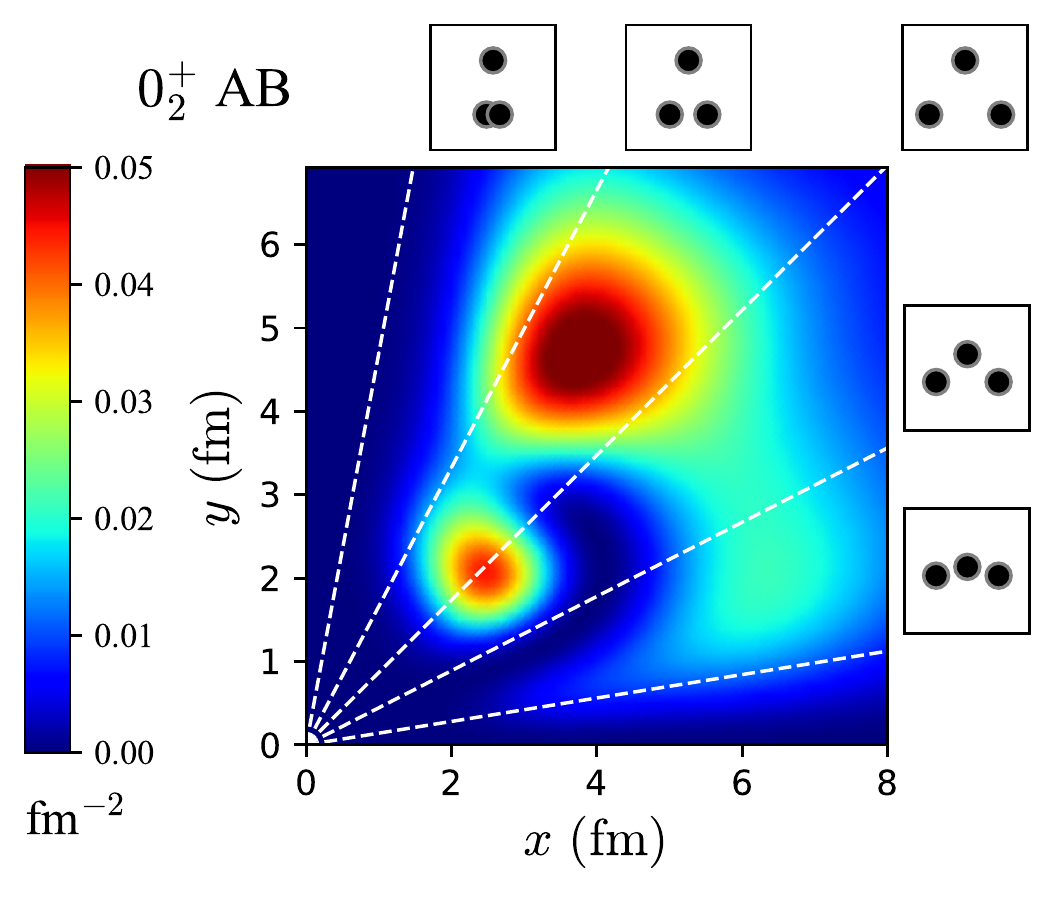}
    \includegraphics[width=0.5\hsize]{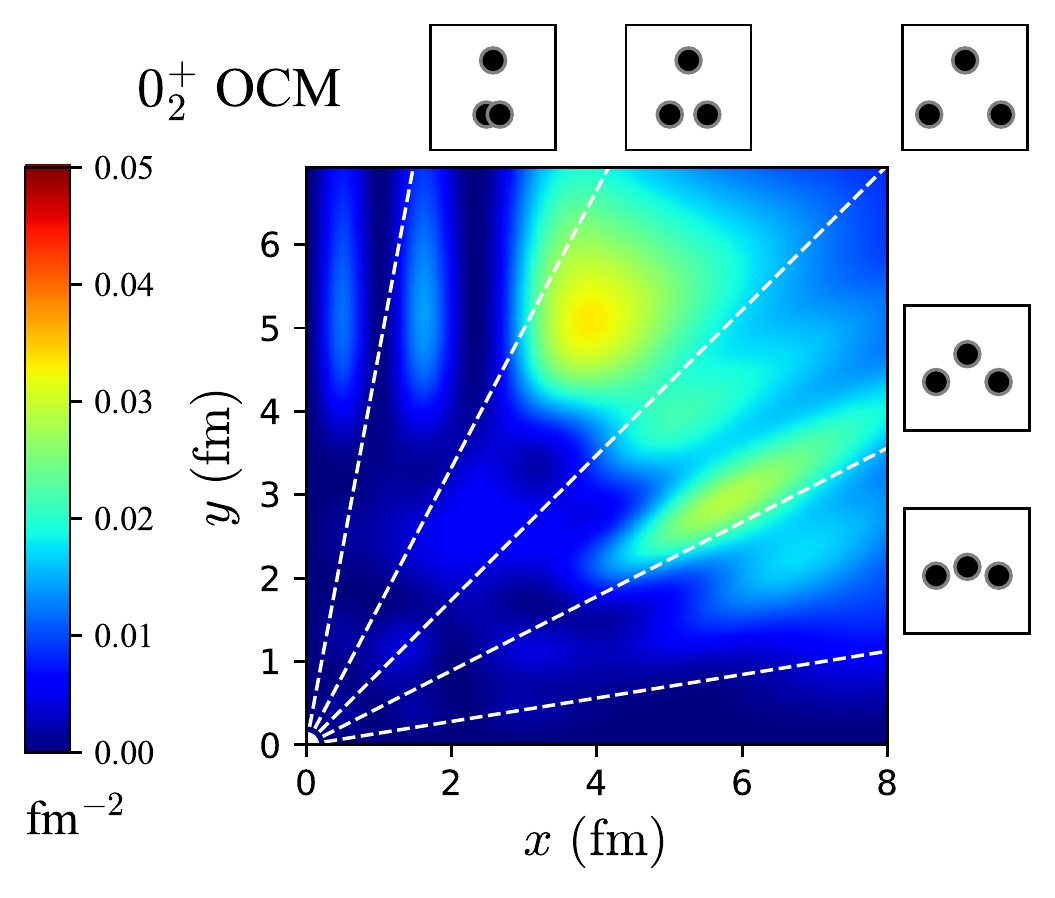}    
    \caption{Two-body density distributions
      for $0_1^{+}$ and $0_2^+$ with the AB and OCM.
      The white dashed lines are plotted as guide for
      some specific geometric configurations that are illustrated
      in the boxes at the end of the lines, e.g.,
      the diagonal line corresponds to the equilateral triangle configuration.}
  \label{fig:tbd}
\end{figure}

To discuss the configurations of three-$\alpha$ particles,
we calculate the two-body density distributions (TBD), 
$\rho(x,y)=
  \left<\Psi_{3\alpha}\left|\delta(|\bm{x}_1|-x)\delta(|\bm{x}_2|-y)
  \right|\Psi_{3\alpha}\right>,$
where $\bm{x}_1=\bm{r}_1-\bm{r}_2$ and $\bm{x}_2=(\bm{r}_1+\bm{r}_2)/2-\bm{r}_3$.
Note $\int_0^\infty dx\int_0^\infty dy\,\rho(x,y)=1$.
Figure~\ref{fig:tbd} plots the TBD
of $0_1^+$ and $0_2^+$ obtained by the AB and OCM. 
For $0_{1}^{+}$, the TBD for the AB
simply has only one peak at the equilateral triangle shape,
while for the OCM, the TBD shows other peak structures
in regions of acute- and obtuse-isosceles triangles.
In the OCM, several small peaks appear in the internal region, which
come from the orthogonality condition on the Pauli forbidden states
imposed in the solution of the OCM. 
On the other hand, for the AB, no peak is found
due to the repulsive component contained in the AB potential.
For $0_{2}^{+}$, the TBD is much extended
and shows more complicated structures.
For the AB, the TBD has predominantly two peaks
in the equilateral and acute-isosceles triangle regions.
The highest peak is located at
the acute-isosceles triangles which implies
the $^{8}\mathrm{Be}+\alpha$ configuration.
For the OCM, though there are several peaks in the internal region
due to the orthogonality condition,
the dominant configurations are isosceles triangles,
almost equal contributions from
the acute- and obtuse-isosceles triangles.

\begin{figure}[h]
  \begin{center}
    \includegraphics[width=\hsize]{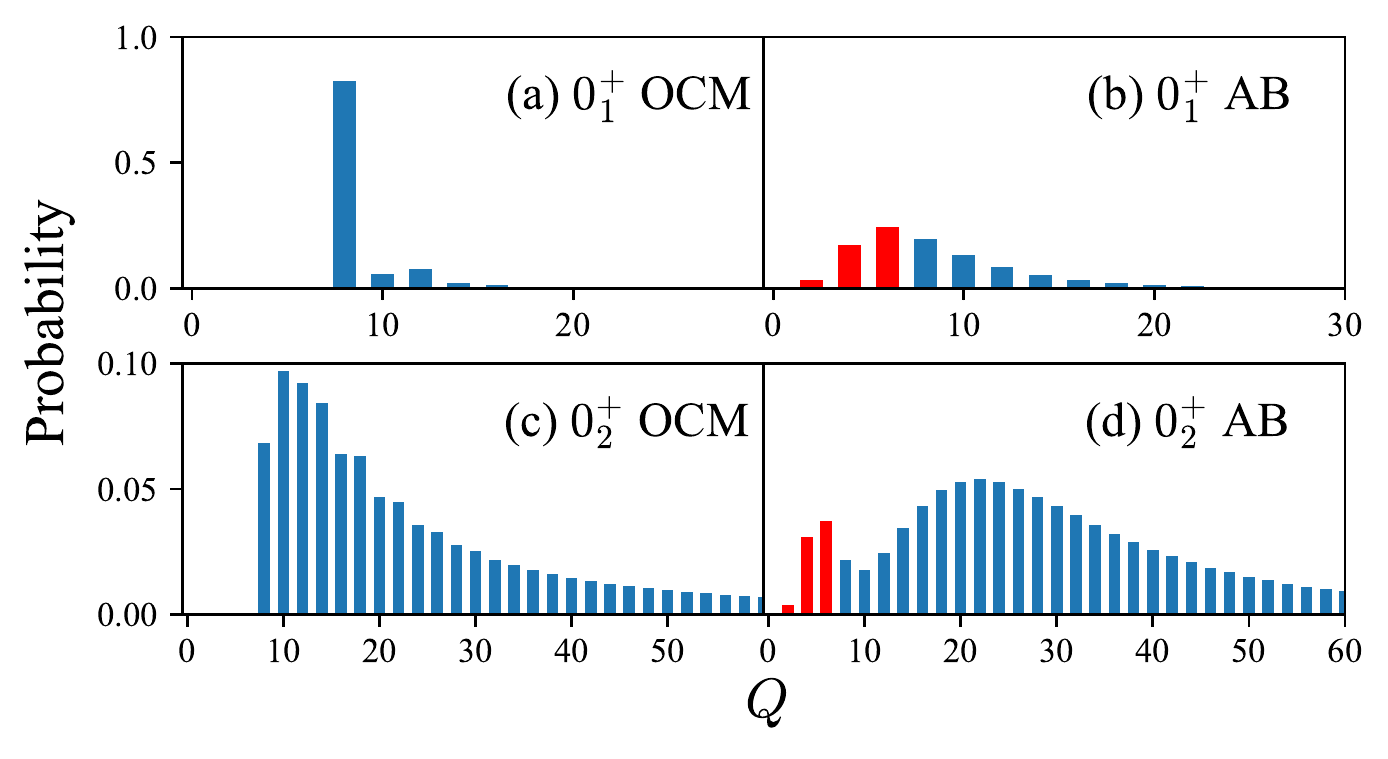}        
  \end{center}
  \caption{Occupation probabilities of
    the harmonic oscillator quanta $Q\hbar \omega$
    for (a), (b) $0_{1}^{+}$ and (c), (d) $0_2^+$.
    The oscillator energy $\hbar \omega$ is set to be
    $22~\mathrm{MeV}$, consistently with the size parameter
    of the forbidden states used in the OCM.    
    The states that are forbidden in the OCM ($Q=0$--6) are colored in red.}
  \label{fig:q}
 \end{figure}

Here we compare the wave function components for those two different
$\alpha$ cluster models. 
Since the two models treat the Pauli principle differently,
it is interesting to see the occupation probability
of the harmonic oscillator quanta in the wave functions~\cite{Suzuki96,hs14}.
Figures~\ref{fig:q} (a) and (b) plot the occupation probabilities for $0_1^+$
with the OCM and AB, respectively.
According to the SU(3) limit of the three-$\alpha$ clusters~\cite{Smirnov74},
the allowed state is $Q\geq 8$.
In fact, for the OCM no forbidden state component with $Q<8$ and 
the $Q=8$ state is dominant, which is
consistent with the result given in Ref.~\cite{ys05}.
In contrast, for the AB, the distribution shows quite different
behavior: It is peaked at $Q=6$
and spreads, having contributions of $Q<8$.
Figures~\ref{fig:q} (c) and (d) show
the occupation probabilities for $0_{2}^{+}$.
Similarly to the $0_1^+$ case,
no forbidden state component with $Q<8$ is included in the OCM,
while they are found in the AB.
Though there are some quantitative differences in the peak positions,
the global behavior of the occupation probabilities
are similar reflecting the characteristic behavior of well-developed
cluster states~\cite{Suzuki96,Neff12,hs14,Horiuchi14}.

To corroborate the discussion, we list in Tab.~\ref{tab:rpw}
the probability of finding partial wave $(l)$ components between
the two-$\alpha$ particles in the three-$\alpha$ system,
$P_{l}$.
In the OCM, $l=0,2,$ and 4 components are almost equal for
$0_{1}^{+}$. This again confirms the SU(3) character
of the wave function~\cite{ys05}.
Reminding $Q=2n+l=8$ with $n$ being the number of nodes,
the dominant $Q=8$ component induces, e.g,. four nodes with $l=0$.
For the AB, differently from the OCM result, the $0_1^+$ state is $l=0$ dominance.
For $0^+_2$, $P_{l=0}$ is dominant for both the OCM and AB,
which is consistent with the $\alpha$ condensed picture~\cite{ys05,Yamada12}
in which the three-$\alpha$ particles are occupied
in the lowest orbit as bosons.

\begin{table}[h]
  \caption{Probability of finding partial wave $(l)$ components
    for $0_{1}^+$ and $0_2^+$ with the OCM and AB.
    The values in parenthesis are the spectroscopic factor
  of $^8{\rm Be}+\alpha$.}
  \label{tab:rpw}
  \begin{center}
    \begin{tabular}{ccccc} \hline
      &&$P_{l=0}$ &$P_{l=2}$ & $P_{l=4}$ \\ \hline \hline
      \multirow{2}{*}{$0_{1}^{+}$}&OCM& $0.295$ (0.195) & 0.348 & 0.313 \\
      &AB& $0.704$ (0.448) & 0.294 & 0.001
      \\ \hline
      \multirow{2}{*}{$0_{2}^{+}$} &OCM& $0.693$ (0.693) & 0.141 & 0.087 \\
      &AB& $0.886$ (0.729) & 0.108 & 0.003
       \\ \hline
    \end{tabular}
  \end{center}
\end{table}

\begin{figure}[htbp]
  \includegraphics[width=0.5\hsize]{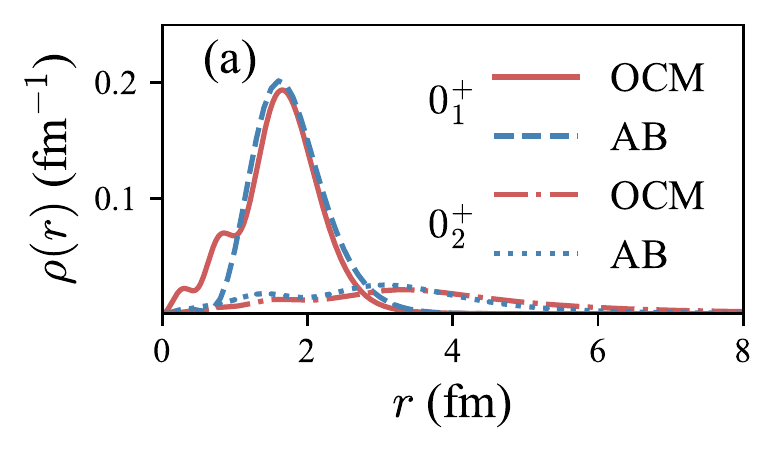}
  \includegraphics[width=0.5\hsize]{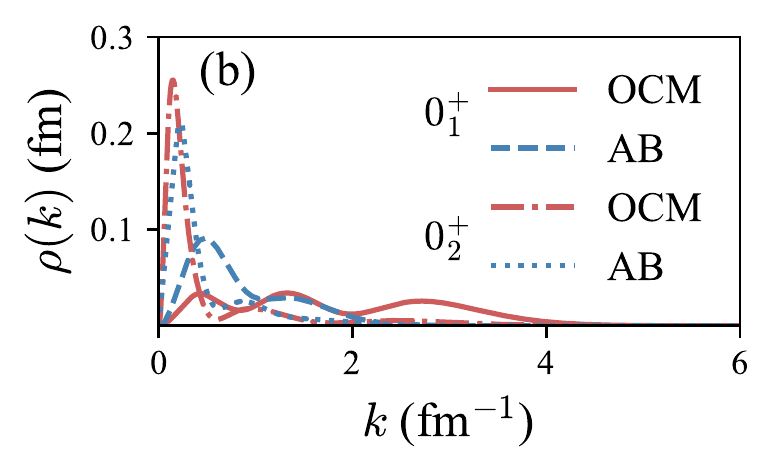}  
  \caption{One-body density distributions in (a) coordinate and (b) 
   momentum space for $0_{1}^{+}$ and $0_{2}^{+}$ with the AB and OCM.}
  \label{fig:odmd}
\end{figure}

These differences are also well reflected
in the one-body density distribution (OBD) defined as
$\rho(r)=\braket{\Psi_{3\alpha}|\delta(|\bm{r}_{1}-\bm{r}_{\rm cm}|-r)|\Psi_{3\alpha}}$,
where $\bm{r}_{\rm cm}$ denotes the center-of-mass coordinate.
Figure~\ref{fig:odmd} (a) plots the OBD in the coordinate space. 
For $0_1^+$ with the OCM, as we see in the TBD,
the OBD also exhibits some nodes in the internal region due
to the orthogonal condition on the Pauli forbidden states,
while no such node exists in the AB result.
For $0_2^+$, the OBDs for the OCM and the AB show similar nodal behavior
due to the contributions of various $Q$ components shown
in Fig.~\ref{fig:q}. Figure~\ref{fig:odmd} (b)
shows the OBD in the momentum space or the momentum distribution,
defined by $\rho(k)=\braket{\Psi_{3\alpha}|\delta(|\bm{k}_{i}-\bm{k}_{\rm cm}|-k)|\Psi_{3\alpha}}$,
where $\bm{k}_1$ is the conjugate momentum of $\bm{r}_1$ and $\bm{k}_{\rm cm}=0$. 
For $0_{1}^{+}$, the momentum distribution of
the OCM has three peaks with almost equal height.
This represents that the three-$\alpha$ particles
occupy in the single $\alpha$ orbits showing the shell-like structure
with $l=0, 2,$ and $4$ in the SU(3) model~\cite{ys05}.
For the AB, the momentum distribution of $0_{1}^{+}$ has a two-peak structure
and the tallest peak is placed at $k\approx 0.5~\mathrm{fm}^{-1}$,
although they are forbidden in the OCM (See Fig.~\ref{fig:q}).
For $0_2^+$, two distributions are quite similar
and their amplitudes are concentrated at low-momentum regions, showing
the characteristics of the BEC~\cite{THSR,Yamada12}.

\begin{figure}[htbp]
  \begin{center}
    \includegraphics{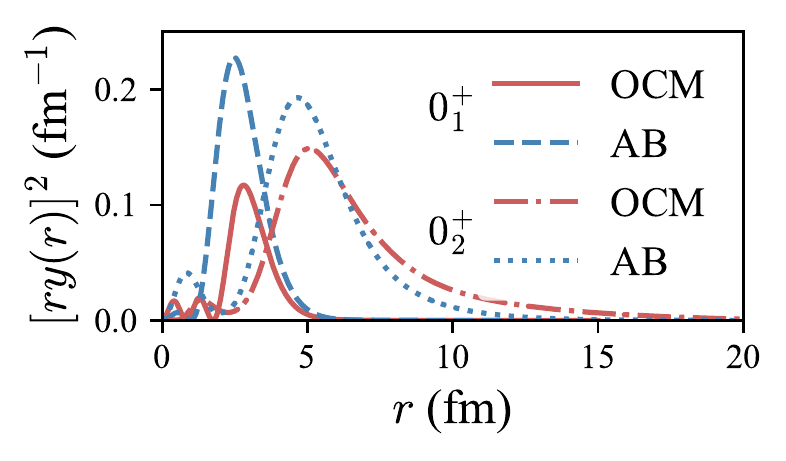}
  \end{center}
  \caption{Spectroscopic amplitudes of $^{8}\mathrm{Be} + \alpha$
    for $0_1^+$ and $0_2^+$ with the OCM and AB.}
  \label{fig:sa}
\end{figure}

Finally, we analyse the component of the $^{8}{\rm Be}+\alpha$
configuration in the three-$\alpha$ wave functions
to clarify its impact on the geometrical configuration.
The spectroscopic amplitude (SA), $y(r)$, is defined by an overlap amplitude
between $\Psi_{2\alpha}$ and $\Psi_{3\alpha}$ as a function
of the relative distance from the center-of-mass of $\Psi_{2\alpha}$
to the other $\alpha$,
where the $\Psi_{2\alpha}$ is the ground-state wave function of $^{8}$Be
with $J^\pi=0^+$. $\int_0^\infty [ry(r)]^2dr=S$ holds
with the spectroscopic factor (SF),
$S=\left|\sqrt\frac{3!}{2!1!}\braket{\Psi_{2\alpha}Y_{00}(\bm{x}_2)|\Psi_{3\alpha}}\right|^{2}$.
Note that SF is a subset of $P_{l=0}$
and is listed in parentheses of Tab.~\ref{tab:rpw}.
The calculated $S$ of $0_{1}^{+}$ for the OCM is small
as $P_{l=0}$ is small, while
large $^8{\rm Be}+\alpha$ component is found for the AB.
For $0_{2}^{+}$ both models give similar
SFs $\approx 0.7$,
showing well-developed $^{8}\mathrm{Be} + \alpha$ configurations.
Figure~\ref{fig:sa} shows $[ry(r)]^2$.
For $0_{1}^{+}$,
more nodes appear for the OCM in the internal region as expected.
Due to the orthogonality condition,
the peak is somewhat shifted to outer regions for the OCM.
For $0_{2}^{+}$, the positions of these peaks are located at $\approx 5$ fm.
Note that the calculated rms distances of $^{8}$Be are
6.36 and 5.93 fm for the AB and OCM, respectively, which are
larger than the respective peak positions of the SA.
Reminding that the $S$ value is quite large for $0_2^+$,
this $^8{\rm Be}+\alpha$ correlation
originates the peak structures of the isosceles triangle shape
in $0_2^+$ shown in Fig.~\ref{fig:tbd}.
We see notable difference between the results of the AB and OCM in
the asymptotic regions,
reflecting the rms radii obtained by these different models,
3.39 and 4.58 fm for the AB and OCM, respectively.
A careful investigation including comparison to available experimental
data is necessary to clarify the dominant $\alpha$-cluster configurations
in the spectrum of $^{12}$C.

\vspace{-0.3cm}
\section{Summary and future perspectives}
\label{sec:6}

To extract geometric configurations of three-$\alpha$
particles in the ground and first excited $0^+$ states of $^{12}$C,
we have performed precise three-$\alpha$ calculations
using two types of potential models, the orthogonality condition model (OCM)
and the shallow potential model (AB),
which produce different internal structure of the wave functions.
Analyzing the wave functions in detail,
we show that the $^{8}{\rm Be}+\alpha$ configuration plays a crucial
role to form isosceles triangle shape in the first excited $0^+$ state.
The notable differences in the wave functions with the OCM and AB
can possibly be distinguished by physical observables.
A comparison with experiment, e.g.,
the decay properties and electromagnetic transitions
including other $J^\pi$ state is necessary
to verify these adopted models for
establishing the three-$\alpha$ configurations in
the spectrum of $^{12}$C.


\vspace{-0.3cm}
\begin{thebibliography}{99}
\bibitem{h54} F. Hoyle,
Astrophys. J. Suppl. \textbf{1}, 121 (1954)
\bibitem{THSR}A. Tohsaki, H. Horiuchi, P. Schuck, G. R\"{o}pke, 
Phys. Rev. Lett. {\bf 87}, 192501 (2001)
\bibitem{f19} L. Fortunato,
Phys. Rev. C \textbf{99}, 031302(R) (2019)
\bibitem{vcfl20} A. Vitturi, J. Casal, L. Fortunato, E.~G. Lanza,
Phys. Rev. C \textbf{101}, 014315 (2020)
\bibitem{OCM} S. Saito, Prog. Theor. Phys. {\bf 40}  893 (1968);
  {\bf 41} 705 (1969); Suppl. {\bf 62}, 11 (1977) 
\bibitem{ab65} S. Ali, A. R. Bodmer, Nucl. Phys. \textbf{80}, 99-112 (1966)
\bibitem{Pinilla11} E.C.Pinilla,D.Baye,P.Descouvemont,W.Horiuchi,Y.Suzuki, Nucl. Phys. {\bf A 865}, 43 (2011)
\bibitem{Arai18} T. Arai, W. Horiuchi, D. Baye,
  Nucl. Phys. {\bf A 977}, 82 (2018)
\bibitem{kk05} C. Kurokawa, K. Kat\={o},
Phys. Rev. C \textbf{71}, 021301 (2005)
\bibitem{kk07} C. Kurokawa, K. Kat\={o},
Nucl. Phys. A \textbf{792}, 87-101 (2007)
\bibitem{Kukulin78} V. I. Kukulin, V. N. Pomenertsev, Ann. Phys. (N. Y.) {\bf 111}, 330 (1978)
\bibitem{Fedorov96}
  D.~V. Fedorov, A.~S. Jensen, Phys.\ Lett.\ {\bf B 389}, 631 (1996)  
\bibitem{i13} S. Ishikawa,
Phys. Rev. C \textbf{87}, 055804 (2013)
\bibitem{i14} S. Ishikawa,
Phys. Rev. C \textbf{90}, 061604(R) (2014)
\bibitem{nnt13} N. B. Nguyen, F. M. Nunes, I. J. Thompson,
Phys. Rev. C \textbf{87}, 054615 (2013)
\bibitem{vs95} K. Varga, Y. Suzuki,
Phys. Rev. C \textbf{52}, 2885 (1995)
\bibitem{SVM} Y. Suzuki, K. Varga,  {\it Stochastic Variational Approach to Quantum-Mechanical
  Few-Body Problems}, Lecture Notes in Physics, Vol. m54 (Springer, Berlin, 1998)
\bibitem{LNP20} Lai Hnin Phyu, H. Moriya, W. Horiuchi, K. Iida, K. Noda,
  M.~T. Yamashita, Prog. Theor. Exp. Phys. {\bf 2020}, 093D01 (2020)
  \bibitem{Suzuki08} Y. Suzuki, W. Horiuchi, M. Orabi, K. Arai,
  Few-Body Syst. {\bf 42}, 33 (2008)
\bibitem{Mitroy13} J. Mitroy, S. Bubin, W. Horiuchi, Y. Suzuki, 
L. Adamowicz, W. Cencek, K. Szalewicz, J. Komasa, D. Blume, K. Varga,
Rev. Mod. Phys. {\bf 85}, 693-749 (2013)
\bibitem{Suzuki17} Y. Suzuki, W. Horiuchi,
  {\it Emergent Phenomena in Atomic Nuclei from Large-scale Modeling: A Symmetry-Guided Perspective} (World Scientific, Singapore, 2017),
  Chap. 7,  pp. 199-227
\bibitem{ofkh13} S. Ohtsubo, Y. Fukushima, M. Kamimura, E. Hiyama,
Prog. Theor. Exp. Phys. \textbf{2013}, 073D02 (2013)
\bibitem{Suzuki96}
Y. Suzuki, K. Arai, Y. Ogawa, K. Varga, Phys. Rev. C {\bf 54}, 2073 (1996)
\bibitem{hs14} W. Horiuchi, Y. Suzuki,
Phys. Rev. C \textbf{90}, 034001 (2014)
\bibitem{Smirnov74} Yu. F. Smirnov, I.T. Obukhovsky, Yu. M. Tchuvil'sky, V. G. Neudatchin, Nucl. Phys. A {\bf 235} 289 (1974)
\bibitem{ys05} T. Yamada, P. Schuck,
Eur. Phys.J. A \textbf{26}, 185-199 (2005)
\bibitem{Horiuchi14} W. Horiuchi, Y. Suzuki,
  Phys. Rev. C {\bf 89}, 011304(R) (2014)
\bibitem{Neff12} T. Neff, J. Phys. Conf. Ser. {\bf 403}, 012028 (2012)
\bibitem{Yamada12} T. Yamada, Y. Funaki, H. Horiuchi, G. R\"{o}pke, P, Schuck, A. Tohsaki,
in {\it Lecture Notes Physics Cluster Nuclei}, edited by C. Beck (Springer, Berlin, Heidelberg, 2012),
Vol.~2, Chap.~5, pp. 229-298
\end{thebibliography}
\end{document}